\documentstyle[epsfig]{elsart}
\begin{document}

\newcommand{\re}{\mathop{\mathrm{Re}}}
\newcommand{\im}{\mathop{\mathrm{Im}}}
\newcommand{\I}{\mathop{\mathrm{i}}}
\newcommand{\D}{\mathop{\mathrm{d}}}
\newcommand{\E}{\mathop{\mathrm{e}}}

\def\lambar{\lambda \hspace*{-5pt}{\rule [5pt]{4pt}{0.3pt}} \hspace*{1pt}}

{\Large  DESY 04-012}

{\Large  January 2004}

\bigskip

\begin{frontmatter}

\journal{Optics Communications}

\date{}

\title{
Design formulas for short-wavelength FELs}

\author{E.L.~Saldin},
\author{E.A.~Schneidmiller},
\author{M.V.~Yurkov}

\address{Deutsches Elektronen-Synchrotron (DESY),
Notkestrasse 85, D-22607 Hamburg, Germany}

\begin{abstract}

Simple formulas for optimization of VUV and X-ray SASE FELs are presented.
The FEL gain length and the optimal beta-function are explicitly expressed in
terms of the electron beam and undulator parameters. The FEL saturation
length is estimated taking into account quantum diffusion in the undulator.
Examples of the FEL optimization are given.
Parameters of a SASE FEL, operating at the Compton wavelength,
are suggested.

\end{abstract}

\end{frontmatter}

\baselineskip 20pt

\clearpage

\section{Introduction}

Successful operation of the VUV (vacuum ultraviolet) FEL (free electron laser)
at the TESLA Test Facility at DESY \cite{109nm,epj}, based on
SASE (self-amplified spontaneous emission) principle \cite{dks},
has stimulated a rapidly growing interest in the development of
VUV and X-ray FELs. A number of projects (see, for instance,
\cite{ttf2,bessy,japan,lcls,tesla})
are now at different stages of design and construction.

At the first stage of a SASE FEL design one looks for the dependence of the FEL saturation
length on the wavelength, electron beam parameters, undulator parameters, and beta-function.
Usually the parameters are optimized for the shortest design wavelength since the saturation length
is the largest in this case.
The saturation length is proportional to the gain length (e-folding length) of the
fundamental transverse mode (see \cite{book} for more details). The gain length
can be found by the solution of the FEL eigenvalue equation.

The eigenvalue equation for a high-gain FEL,
including diffraction of radiation, emittance, and energy spread, was
derived in \cite{kim,yu88}. There exist approximate solutions \cite{yu90,chin92}
of this equation. The exact solution was
presented in \cite{ming} as well as an approximate solution (with a limited validity range).
The latter solution was fitted \cite{ming} using 3 dimensionless
groups of parameters, and 19 fitting coefficients.
An approximate solution, that fits the exact solution in the entire parameter space with
high accuracy (better that 1 \%), was presented in \cite{emit1}. A numerical algorithm for
finding this approximate solution is very fast and robust. It was used to obtain the
main results of this paper.

In this paper we present the explicit, simple and
rather accurate dependencies
of the FEL gain length on the beam and undulator parameters.
Our formulas are not universal, but they provide a good accuracy (better than
5 \% for the gain length) in a typical parameter range of
VUV and X-ray FELs. We present
the formulas without derivation since they
were not derived analytically. In some sense the parametric dependencies were guessed,
and then the fitting coefficients were found from
the solution of the eigenvalue equation. For instance, we used only 2 fitting coefficients
for the gain length with the optimized beta-function.
Our formulas allow one to quickly estimate FEL saturation length, including the
principal effect of energy diffusion in the undulator due to quantum
fluctuations of the undulator radiation. In addition, we present two
practical examples of using our design formulas: optimization of SASE FEL with negligible
energy spread, and the limitation on SASE FEL wavelength taking into account quantum
diffusion. In particular, we suggest for the first time the set of parameters for
a SASE FEL operating at the Compton wavelength.

\section{Gain length for the optimized beta-function}

Let us consider an axisymmetric electron beam with a current $I$, and a Gaussian distribution
in transverse phase space and in energy \cite{ming,emit1}. The focusing structure in the
undulator is
a superposition of the natural undulator focusing and an external alternating-gradient
focusing. The eigenvalue equation \cite{ming,emit1} is valid under the following
condition \cite{emit1}:

\begin{displaymath}
\frac{L_{\mathrm{f}}}{2\pi \beta} \ll \mathrm{min} \left(
1, \, \frac{\lambda_r}{2 \pi \epsilon} \right)
\end{displaymath}

\noindent where $L_{\mathrm{f}}$ is the period of the external focusing structure,
$\beta$ is an average beta-function, $\epsilon$ is the rms emittance of the electron beam,
and $\lambda_r$ is the FEL resonant wavelength. The resonance condition is written as:

\begin{equation}
\lambda_r = \frac{\lambda_{\mathrm{w}}(1+K^2)}{2 \gamma^2} \ .
\label{resonance}
\end{equation}

\noindent Here $\lambda_{\mathrm{w}}$ is the undulator period, $\gamma$ is relativistic
factor, and $K$ is the rms undulator parameter:

\begin{equation}
K = 0.934 \ \lambda_{\mathrm{w}} [{\mathrm{cm}}] \ B_{\mathrm{rms}} [{\mathrm{T}}] \ ,
\label{k-rms}
\end{equation}

\noindent $B_{\mathrm{rms}}$ being the rms undulator field.

In what follows we assume that the beta-function is optimized so that the FEL gain length
takes the minimal value for given wavelength, beam and undulator parameters.
Under this condition the solution of the eigenvalue equation for the field gain
length\footnote{There is also a notion of the power gain length which is twice shorter.}
can be approximated as follows:

\begin{equation}
L_g \simeq L_{g0} \ (1+\delta) \ ,
\label{lg}
\end{equation}

\noindent where

\begin{equation}
L_{g0} = 1.67 \left(\frac{I_A}{I} \right)^{1/2} \frac{(\epsilon_n \lambda_{\mathrm{w}})^{5/6}}
{\lambda_r^{2/3}} \ \frac{(1+K^2)^{1/3}}{K A_{JJ}} \ ,
\label{lg0}
\end{equation}

\noindent and

\begin{equation}
\delta = 131 \ \frac{I_A}{I} \ \frac{\epsilon_n^{5/4}}
{\lambda_r^{1/8} \lambda_{\mathrm{w}}^{9/8}} \ \frac{\sigma_{\gamma}^2}{(K A_{JJ})^2 (1+K^2)^{1/8}} \ .
\label{delta}
\end{equation}

\noindent The following notations are introduced here: $I_A = 17$ kA is the Alfven current,
$\epsilon_n = \gamma \epsilon$ is the rms normalized emittance,
$\sigma_{\gamma}=\sigma_{_{\cal E}}/mc^2$ is the rms energy spread
(in units of the rest energy), $A_{JJ} = 1$ for a helical
undulator and $A_{JJ} = J_0(K^2/2(1+K^2))-J_1(K^2/2(1+K^2))$ for a planar undulator,
$J_0$ and $J_1$ are the Bessel functions of the first kind.

The formula (\ref{lg}) provides an accuracy better than 5 \% in the domain of
parameters defined as follows

\begin{equation}
1 < \frac{2 \pi \epsilon}{\lambda_r} < 5
\label{emit-lam-lim}
\end{equation}
\begin{equation}
\delta < 2.5 \ \left\{ 1-
\exp \left[ - \frac{1}{2} \, \left( \frac{2 \pi \epsilon}{\lambda_r}
\right)^2 \right]
\right\}
\label{delta-lim}
\end{equation}

\noindent Note that the condition (\ref{emit-lam-lim}) is usually satisfied in realistic
designs of VUV and X-ray FELs when one does optimization for the shortest wavelength
(defining the total undulator length). The condition (\ref{delta-lim}) is practically
not a limitation. To illustrate the accuracy of the formula (\ref{lg}) we present a
numerical example. The following nominal operating point is chosen: $\lambda_r = 1$ nm,
$\lambda_{\mathrm{w}} = 3$ cm, $K=1$, $I = 2.5$ kA, $\epsilon_n = 2 \ \mu$m,
$\sigma_{_{\cal E}} = 1$ MeV, energy is 2.8 GeV, undulator is planar.
We scan over different parameters
and compare the gain length calculated with formula (\ref{lg}) and by solving the
eigenvalue equation \cite{emit1}. The results are presented in Figs. 1-6.

We also present here an approximate expression for the optimal beta-function (an accuracy
is about 10 \% in the above mentioned parameter range):

\begin{equation}
\beta_{\mathrm{opt}} \simeq 11.2 \left(\frac{I_A}{I} \right)^{1/2} \frac{\epsilon_n^{3/2}
\lambda_{\mathrm{w}}^{1/2}}
{\lambda_r K A_{JJ}} \ (1+8\delta)^{-1/3}
\label{beta}
\end{equation}

\noindent Note that dependence of the gain length on beta-function is rather weak
when $\beta > \beta_{\mathrm{opt}}$.

Finally, let us note that the saturation length cannot be directly found from
the eigenvalue equation. However, with an accuracy 10-20 \%
one can accept the following estimate:

\begin{equation}
L_{\mathrm{sat}} \simeq 10 \ L_{g}
\label{lsat}
\end{equation}

\section{Influence of quantum diffusion in an undulator on saturation length}

Energy spread growth due to the quantum fluctuations of the spontaneous
undulator radiation can be
an important effect \cite{dks,qf-limit} in future SASE FELs.
The rate of the energy diffusion is given by \cite{qf-und}:

\begin{equation}
\frac {d \sigma_{\gamma}^2}{dz} =
\frac{14}{15} \lambar_{\mathrm{c}}r_{\mathrm{e}} \gamma ^4
\kappa _{\mathrm{w}}^3 K^2 F(K) \ ,
\label{eq:energy-diffusion-2}
\end{equation}

\noindent where $\lambar_{\mathrm{c}} = 3.86 \times 10^{-11}$ cm,
$r_{\mathrm{e}}=2.82 \times 10^{-13}$ cm,
$\kappa _{\mathrm{w}} = 2\pi/\lambda_{\mathrm{w}}$, and

\begin{eqnarray}
F(K) & = & 1.42 K + (1 + 1.50 K + 0.95 K^2)^{-1}  \ \ \ \ \ \ \ \ \ \
\mathrm{for} \ helical \ undulator \nonumber \\
F(K) & = & 1.70 K + (1 + 1.88 K + 0.80 K^2)^{-1}  \ \ \ \ \ \ \ \ \ \
\mathrm{for} \ planar \ undulator
\label{fk}
\end{eqnarray}

To estimate the FEL saturation length, we accept the following scheme. First, we neglect
energy diffusion and find a zeroth order approximation to the saturation length from
(\ref{lsat}), (\ref{lg})-(\ref{delta}). Then we calculate an induced energy spread
in the middle of the undulator
from (\ref{eq:energy-diffusion-2}), add it quadratically to the initial energy spread,
and find a new expression for $\delta$. Then, using (\ref{lsat}), (\ref{lg})-(\ref{delta}),
we find the first
approximation to the saturation length. Then we do the next iteration, etc. Finally,
the saturation length can be estimated as

\begin{equation}
L_{\mathrm{sat}} \simeq 10 \ L_{g0} \ \frac{1+\delta}{1-\delta_{q}} \ ,
\label{lsat-q}
\end{equation}

\noindent where

\begin{equation}
\delta_{q} = 5.5\times 10^4
\left(\frac{I_A}{I}\right)^{3/2} \frac{\lambar_{\mathrm{c}}r_{\mathrm{e}}\epsilon_n^2}
{\lambda_r^{11/4} \lambda_{\mathrm{w}}^{5/4}} \ \frac{(1+K^2)^{9/4}F(K)}{K A_{JJ}^3}
\label{delta-q}
\end{equation}

\noindent Note that in the latter formula the powers are somewhat simplified.
Comparing Eqs. (\ref{lsat}) and (\ref{lsat-q}), we can introduce an effective parameter

\begin{equation}
\delta_{\mathrm{eff}} = \frac{\delta+\delta_q}{1-\delta_q} \ ,
\label{delta-eff}
\end{equation}

\noindent which should be used instead of $\delta$ in (\ref{delta-lim}) to check
the applicability range and in (\ref{beta}) to estimate the optimal beta-function.

Although formula (\ref{lsat-q}) is rather crude estimate, it can be used for quick
orientation in the parameter space with {\it a posteriori} check using a numerical
simulation code.

\section{Examples of SASE FEL optimization}

\subsection{Optimized FEL with a negligible energy spread}

Formulas, presented in the previous Sections, can be used for the optimization of
undulator parameters as soon as a specific
type of the undulator is chosen.
We demonstrate such a possibility with the planar NdFeB undulator of which magnetic
field can be described by the following formula \cite{tesla}:

\begin{equation}
B_{\mathrm{max}} [{\mathrm{T}}] = 3.694 \ \exp \left[ -5.068 \frac{g}{\lambda_{\mathrm{w}}} +
1.52 \left( \frac{g}{\lambda_{\mathrm{w}}} \right)^2 \right]
\ \ \ \ \ \ \ \ \  {\mathrm{for}} \ \ \ 0.1 \, < \, g/\lambda_{\mathrm{w}} \, < \, 1 \ ,
\label{bmax}
\end{equation}

\noindent where $g$ is the undulator gap. The rms value of the parameter $K$ is
given by Eq.~(\ref{k-rms}) with $B_{\mathrm{rms}}= B_{\mathrm{max}}/\sqrt{2}$.

We assume that the energy spread effect on the FEL operation can be neglected
($\delta \ , \ \delta_{q} \to 0$). Then, using (\ref{lg}), (\ref{k-rms}) and (\ref{bmax}),
we minimize the gain length for a given undulator gap. The optimal undulator period is
found to be

\begin{equation}
(\lambda_{\mathrm{w}})_{\mathrm{opt}} [{\mathrm{cm}}] \simeq 1 + 2 \, g \, [{\mathrm{cm}}]
\ \ \ \ \ \ \ \ \ \ {\mathrm{for}} \ \ \ g \, > \, 0.5 \ {\mathrm{cm}}
\label{lamw-opt}
\end{equation}

\noindent The optimal value of $K$ is then defined from (\ref{bmax}) and (\ref{k-rms}),
the electron beam energy - from (\ref{resonance}), and
the optimal beta-function - from (\ref{beta}). The
minimal gain length can be expressed (in practical units) as follows:

\begin{equation}
(L_{g})_{\mathrm{min}}[{\mathrm{m}}] \simeq 20 \
\frac{\epsilon_n^{5/6}[{\mathrm{\mu m}}] \,
g^{1/2}[{\mathrm{cm}}]}
{I^{1/2}[{\mathrm{kA}}] \, \lambda_r^{2/3}[{\mathrm{\AA}}]}  \ .
\label{lg-min}
\end{equation}

Using estimate of the saturation length (\ref{lsat}), we find the minimal wavelength at
which SASE FEL can saturate within the given undulator length $L_{\mathrm{w}}$:

\begin{equation}
(\lambda_r)_{\mathrm{min}}[{\mathrm{\AA}}] \simeq 3 \times 10^3 \
\frac{\epsilon_n^{5/4}[{\mathrm{\mu m}}] \,
g^{3/4}[{\mathrm{cm}}]}
{I^{3/4}[{\mathrm{kA}}] \, L_{\mathrm{w}}^{3/2}[{\mathrm{m}}]}
\label{lam-min}
\end{equation}

\subsection{SASE FEL at the Compton wavelength}

Another example is the optimization of sub-Angstrom FELs for which the effect of
quantum diffusion in the undulator can play an important role.
We consider the case when the energy spread is dominated by the quantum diffusion, and
neglect initial energy spread ($\delta \to 0$).
Optimizing undulator period and parameter $K$ in (\ref{lsat-q}),  we get the following
estimate for the minimal wavelength\footnote{One can notice
the difference with more crude estimate presented in \cite{qf-limit}}:

\begin{equation}
(\lambda_r)_{\mathrm{min}}^{\mathrm{q}}[{\mathrm{\AA}}] \simeq
\frac{4 \, \epsilon_n [{\mathrm{\mu m}}]}
{I^{3/5}[{\mathrm{kA}}] \, L_{\mathrm{w}}^{2/5}[{\mathrm{m}}]}
\label{lam-min-qf}
\end{equation}

Note that in some cases the optimal undulator parameters can be impractical.
In any case, the estimate (\ref{lam-min-qf}) gives a lower limit.
The following numerical examples show
that one can be close to this limit with technically feasible undulator parameters.

Let us consider the electron beam parameters (peak current and emittance)
assumed in \cite{lcls2}. One of the examples, considered in \cite{lcls2}, is a
SASE FEL operating at $\lambda_r = 0.28 \, \mathrm{\AA}$ with $I = 5 \, {\mathrm{kA}}$ and
$\epsilon_n = 0.3 \, {\mathrm{\mu m}}$. Another example is even more ambitious:
$\lambda_r = 0.12 \, \mathrm{\AA}$ with $I = 5 \, {\mathrm{kA}}$ and
$\epsilon_n = 0.1 \, {\mathrm{\mu m}}$.

We try to push the wavelength closer to the extreme given by Eq.~(\ref{lam-min-qf}). In our
first example we assume $I = 5 \, {\mathrm{kA}}$ and
$\epsilon_n = 0.3 \, {\mathrm{\mu m}}$. With these parameters the wavelength
$\lambda_r = 0.1 \, \mathrm{\AA}$ can be reached at the electron beam energy 23 GeV
in a planar undulator with $\lambda_{\mathrm{w}} = 2$ cm and $K = 1$
(with the gap $g = 0.7$ cm according
to (\ref{bmax}) and (\ref{k-rms})). The optimal beta-function is about 40 m, and the
saturation length is estimated at 160 m.

The second example is a SASE FEL operating at the Compton wavelength,
$\lambda_r = \lambda_{\mathrm{c}} = 0.0234 \, \mathrm{\AA}$ (photon energy is 0.5 MeV).
We assume the electron beam with $I = 5 \, {\mathrm{kA}}$ and
$\epsilon_n = 0.1 \, {\mathrm{\mu m}}$, the energy is 40 GeV. We choose a helical undulator
with $\lambda_{\mathrm{w}} = $ 2 cm and $K = 0.7$. The optimal beta-function is about 35 m,
and the saturation is reached within 200 m.
Our estimates show that quantum effects, other than
energy diffusion, give small corrections to the classical description and can be
neglected.

\clearpage

\begin{figure}
\hspace*{1cm}
\begin{center}
\epsfig{file=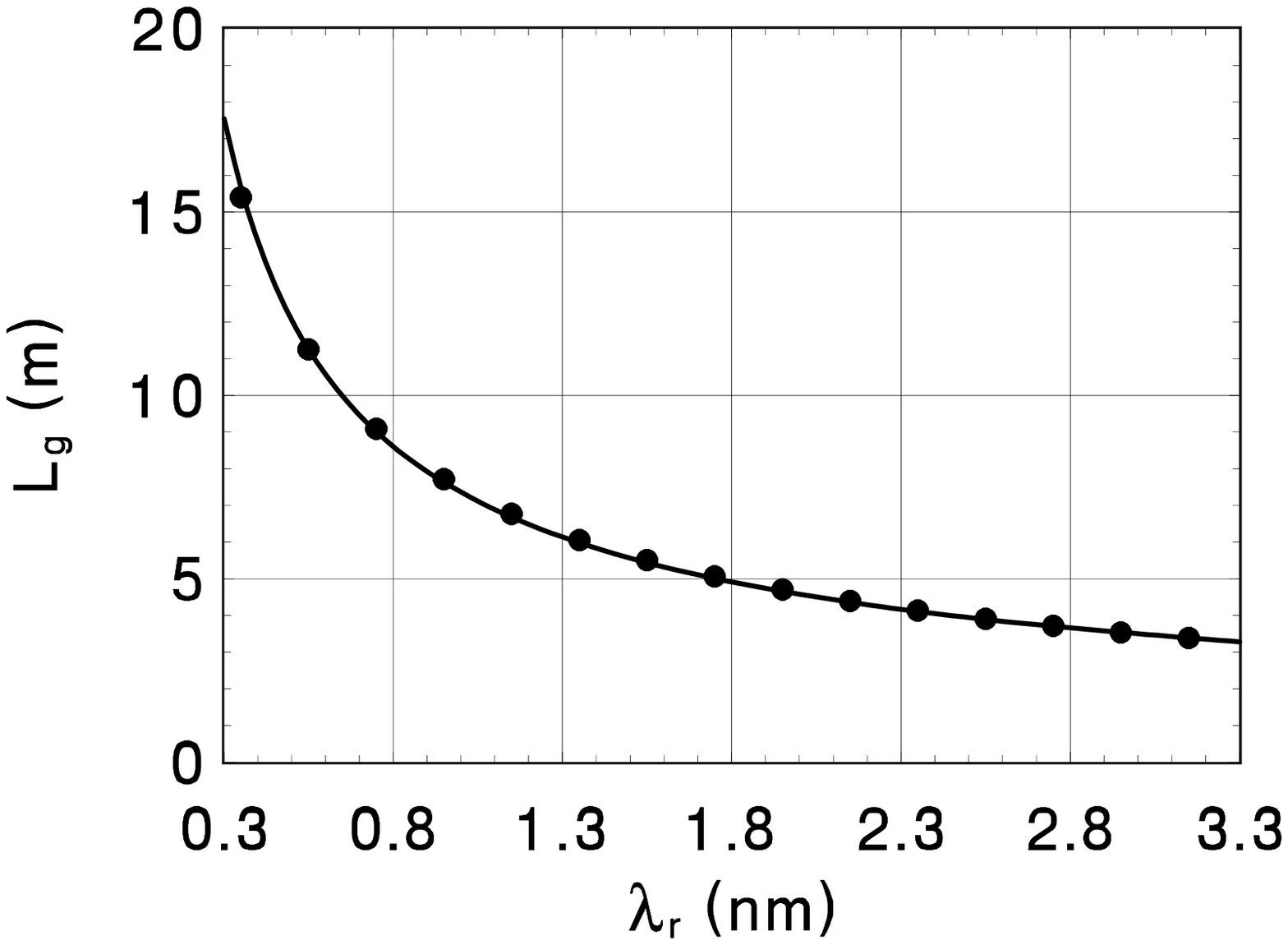,width=0.7\textwidth}
\end{center}
\caption{Gain length versus resonant wavelength for the following set of parameters:
$\lambda_{\mathrm{w}} = 3$ cm, $K=1$, $I = 2.5$ kA, $\epsilon_n = 2 \ \mu$m,
$\sigma_{_{\cal E}} = 1$ MeV. Undulator is planar,
resonance is maintained by tuning electron beam energy,
beta-function is optimized for each case. Line is the solution of the eigenvalue
equation \cite{emit1}, and the circles are calculated using formula (\ref{lg}).}
\hspace*{1cm}
\end{figure}

\bigskip

\begin{figure}
\hspace*{1cm}
\begin{center}
\epsfig{file=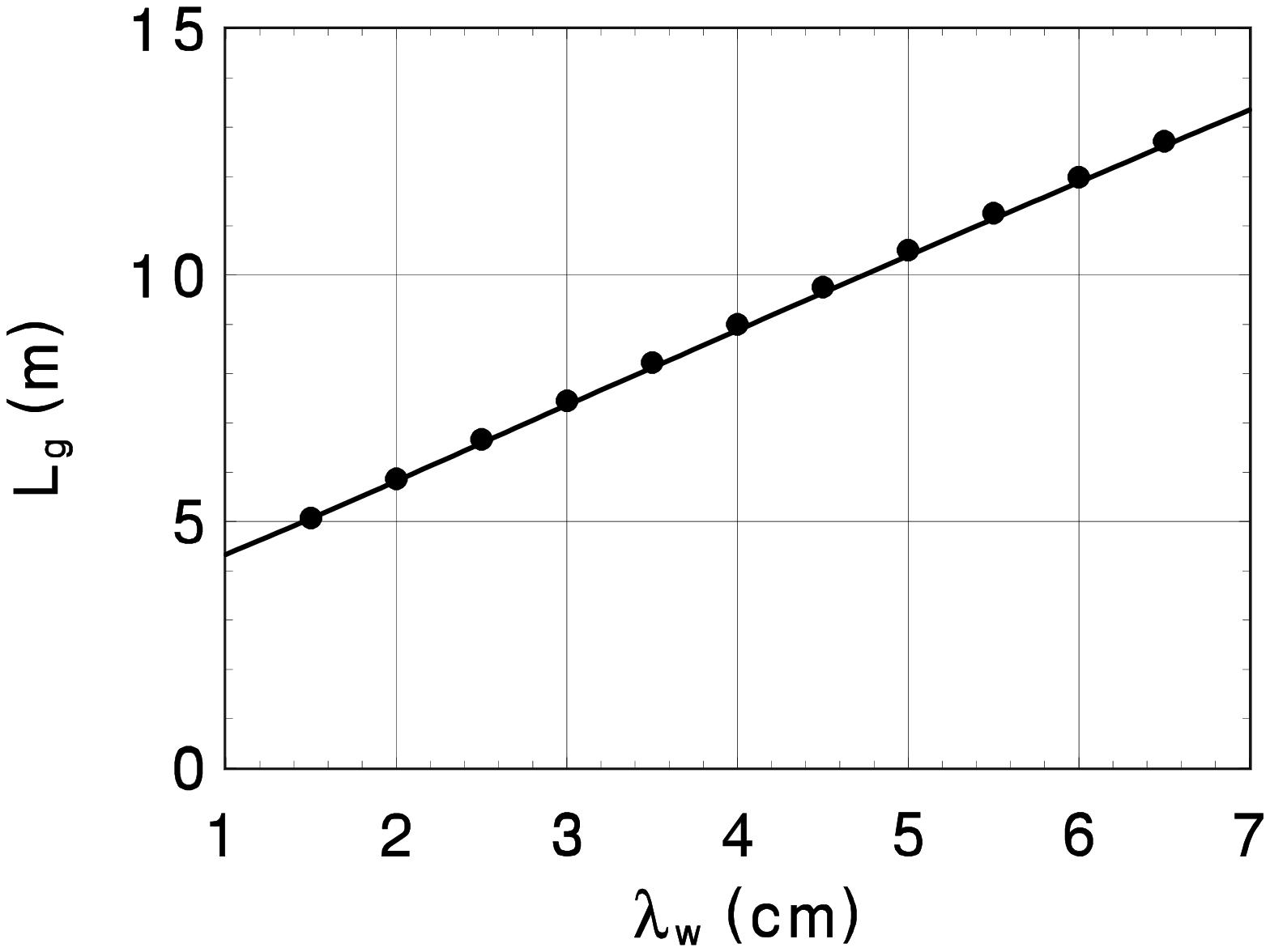,width=0.7\textwidth}
\end{center}
\caption{
Gain length versus undulator period for the following set of parameters:
$\lambda_r = 1$ nm, $K=1$, $I = 2.5$ kA, $\epsilon_n = 2 \ \mu$m,
$\sigma_{_{\cal E}} = 1$ MeV. Undulator is planar,
resonance is maintained by tuning electron beam energy,
beta-function is optimized for each case. Line is the solution of the eigenvalue
equation \cite{emit1}, and the circles are calculated using formula (\ref{lg}).
}
\hspace*{1cm}
\end{figure}

\clearpage

\begin{figure}
\hspace*{1cm}
\begin{center}
\epsfig{file=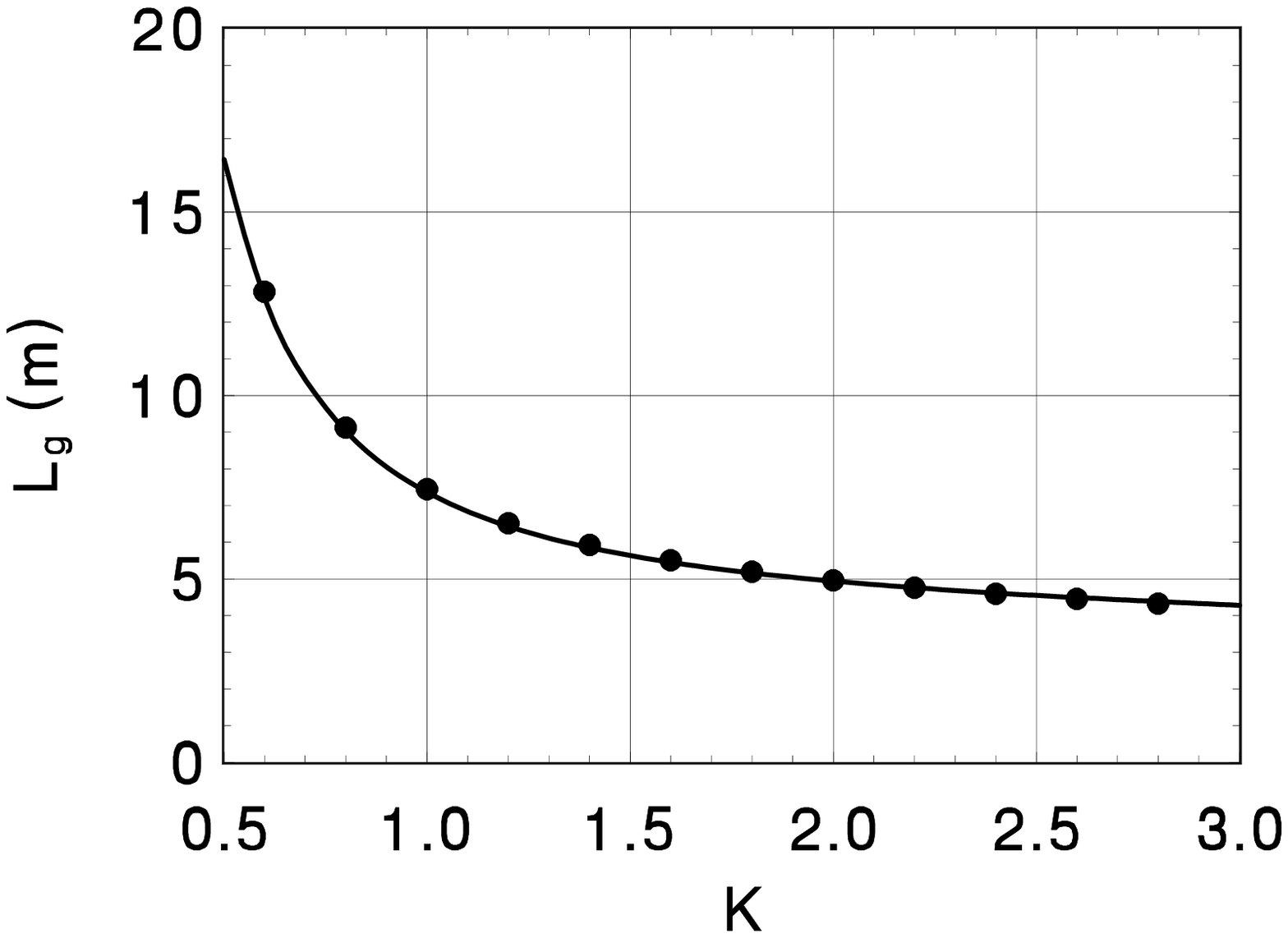,width=0.7\textwidth}
\end{center}
\caption{
Gain length versus undulator parameter K for the following set of parameters:
$\lambda_r = 1$ nm, $\lambda_{\mathrm{w}} = 3$ cm, $I = 2.5$ kA, $\epsilon_n = 2 \ \mu$m,
$\sigma_{_{\cal E}} = 1$ MeV. Undulator is planar,
resonance is maintained by tuning electron beam energy,
beta-function is optimized for each case. Line is the solution of the eigenvalue
equation \cite{emit1}, and the circles are calculated using formula (\ref{lg}).
}
\hspace*{1cm}
\end{figure}

\bigskip

\begin{figure}
\hspace*{1cm}
\begin{center}
\epsfig{file=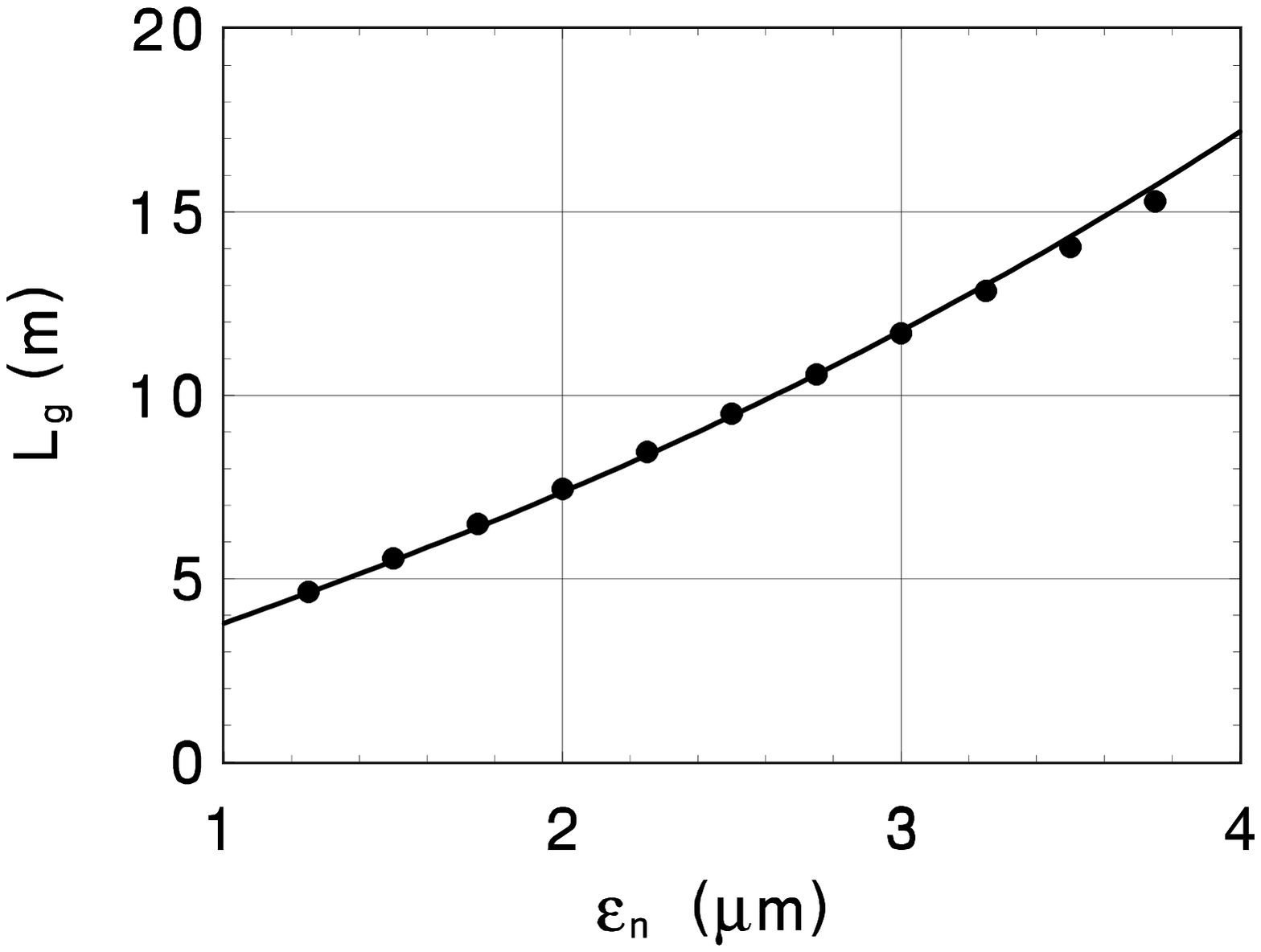,width=0.7\textwidth}
\end{center}
\caption{
Gain length versus normalized emittance for the following set of parameters:
$\lambda_r = 1$ nm, $\lambda_{\mathrm{w}} = 3$ cm, $K = 1$, $I = 2.5$ kA,
$\sigma_{_{\cal E}} = 1$ MeV. Undulator is planar,
beta-function is optimized for each case. Line is the solution of the eigenvalue
equation \cite{emit1}, and the circles are calculated using formula (\ref{lg}).
}
\hspace*{1cm}
\end{figure}

\clearpage

\begin{figure}
\hspace*{1cm}
\begin{center}
\epsfig{file=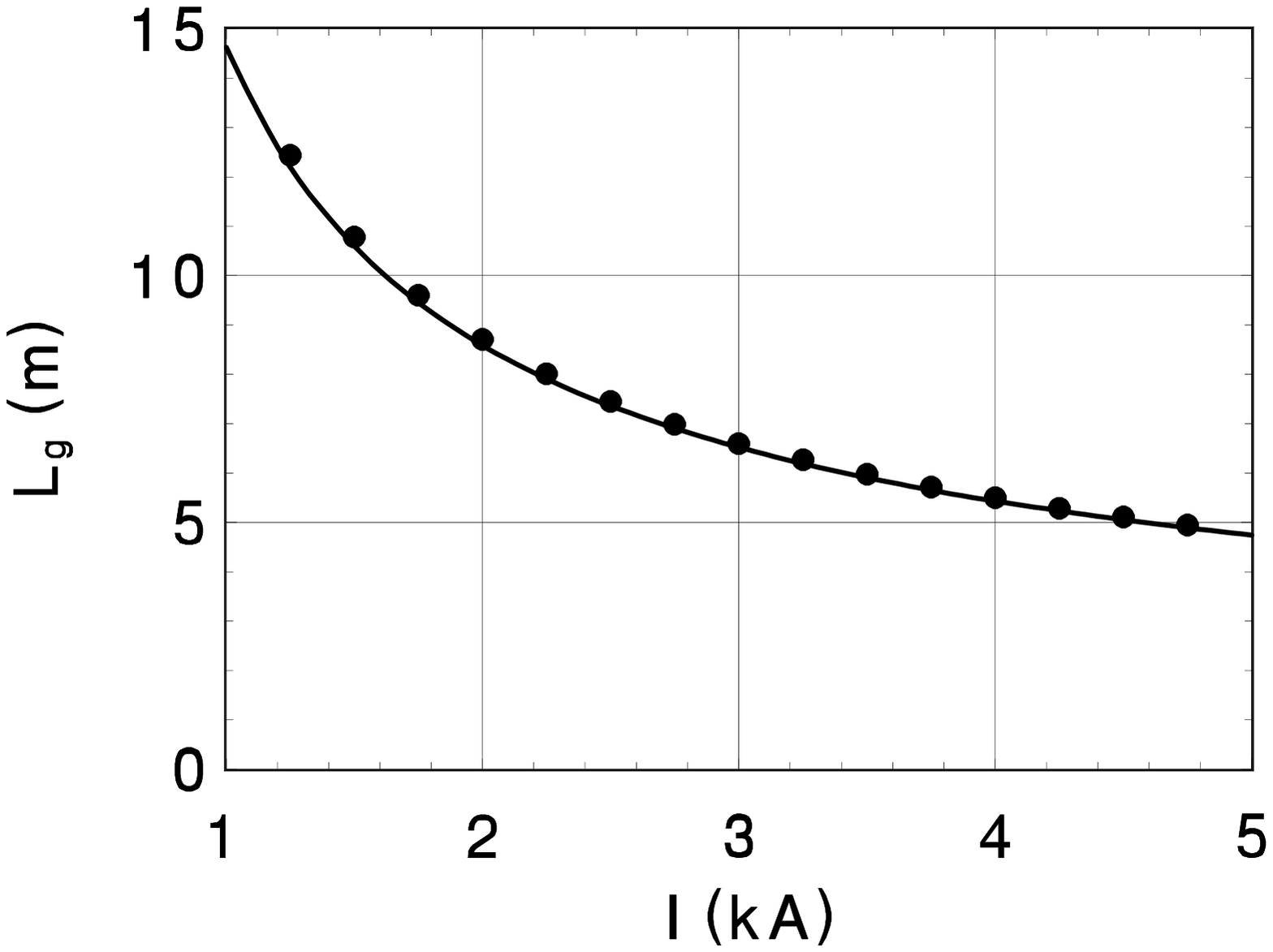,width=0.7\textwidth}
\end{center}
\caption{
Gain length versus current for the following set of parameters:
$\lambda_r = 1$ nm, $\lambda_{\mathrm{w}} = 3$ cm, $K = 1$, $\epsilon_n = 2 \ \mu$m,
$\sigma_{_{\cal E}} = 1$ MeV. Undulator is planar,
beta-function is optimized for each case. Line is the solution of the eigenvalue
equation \cite{emit1}, and the circles are calculated using formula (\ref{lg}).
}
\hspace*{1cm}
\end{figure}

\bigskip

\begin{figure}
\hspace*{1cm}
\begin{center}
\epsfig{file=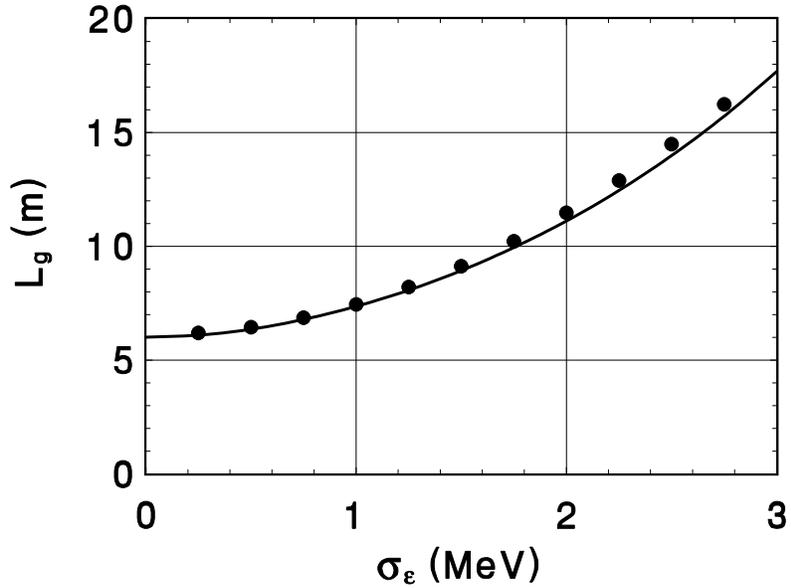,width=0.7\textwidth}
\end{center}
\caption{
Gain length versus energy spread for the following set of parameters:
$\lambda_r = 1$ nm, $\lambda_{\mathrm{w}} = 3$ cm, $K = 1$, $I = 2.5$ kA, $\epsilon_n = 2 \ \mu$m.
Undulator is planar,
beta-function is optimized for each case. Line is the solution of the eigenvalue
equation \cite{emit1}, and the circles are calculated using formula (\ref{lg}).
}
\hspace*{1cm}
\end{figure}

\end{document}